%Paper: astro-ph/9405010
%From: "Arnon Dar, Jaques Goldberg, Michael Rudsky"
%%<PHR19AD@TECHNION.TECHNION.AC.IL>
%Date: Wed, 04 May 94 00:37:43 IST

\input phyzzx

%\input phyzzx

%macropackage=phyzzx
\nopubblock
\PHYSREV
\parindent=0 truecm
\parskip=0 truecm

{\bf
\titlepage
\title{DARK MATTER AND BIG BANG NUCLEOSYNTHESIS}
\author{Arnon Dar, Jacques Goldberg and Michael Rudzsky}
\address{Department of Physics,
Technion - Israel Institute of Technology, Haifa 32000, Israel}}
\baselineskip=16pt plus 2pt
\abstract
Walker, Steigman, Schramm, Olive and Kang (WSSOK) have used the
Standard Big Bang Nucleosynthesis (SBBN)
theory and observed abundances of $^4$He, D+$^3$He and
$^7$Li extrapolated to their primordial values,
to argue that most of the baryons in the
Universe are dark. But, we show here that
the confidence level of the alleged agreement
between the SBBN abundances and those inferred by WSSOK from observations
is less than 0.02\% for any baryon to photon ratio! If, however,  the
highly uncertain WSSOK essentially theoretical upper bound
on the primordial abundance of D+$^3$He
is ignored, then the predicted
abundances of primordial $^4$He, D, $^3$He and $^7$Li are in agreement
with the observations at a confidence level above 70\%,
provided that the cosmic baryon to photon ratio is $\eta=
(1.60\pm 0.1)\times 10^{-10}.$ The predicted primordial D abundance
is $(2.1\pm0.35)\times 10^{-4}$) and may be tested in the future
by measuring absorption line systems in quasar spectra with the Hubble
Space Telescope. The above baryon to photon ratio yields a
baryon mass density, in critical density units, of
$\Omega_{B}\approx 0.0059h^{-2},$  where $h$ is
the Hubble parameter in units of $100~km~s^{-1} Mpc^{-1}.$ This baryon
density is consistent with the observed  mean density of luminous
matter in the Universe (stars, X-Ray emmiting gas, quasar light
absorbing systems), $\Omega_{LUM}\approx 0.0060\pm 0.0027~,$
in particular,
if the most recent estimates, $0.75\leq h\leq 1~,$ are correct. It
does not provide  reliable evidence that most of the
baryons in the Universe are dark. Moreover, since the dynamics
of clusters of galaxies and large scale structures indicate that
$\Omega \gsim 0.15~,$ it does imply  that most of the
matter in the Universe is non baryonic dark matter.
\endpage
{\bf I. INTRODUCTION}

The quantitative agreement between the predictions of the Standard Big
Bang Nucleosynthesis (SBBN) theory (Hayashi 1950; Alpher,
Follin and Herman 1953; Peebles 1966; Wagoner, Fowler and Hoyle 1967;
Wagoner 1969; Wagoner 1973;  Yang et al. 1984; for reviews,
see for instance Weinberg 1972; Schramm and Wagoner 1977; Bosegard and
Steigman 1985; Kolb and Turner 1990; Olive 1991)
and the observed abundances of
$^4$He, D+$^3$He and $^7$Li extrapolated to their primordial values
(see for instance Walker, Schramm, Steigman, Olive and Kang
 1991, hereafter WSSOK) has been
considered as one of the most convincing evidences for the validity
of the Standard Hot Big Bang Model of the Universe (see
for instance Weinberg 1972; Peebles et al. 1991).
The SBBN predictions of the primordial abundances of the light
elements depend essentially on well known nuclear reaction rates
and on three additional parameters,

1) the number of generations of light neutrinos $N_\nu,$

2) the neutron lifetime $\tau_n~,$

3) the ratio of baryons to photons
    in the Universe $\eta\equiv n_{B}/n_\gamma~.$

Recently $N_\nu,$ $\tau_n$ and $n_\gamma$ have also been measured
quite accurately: The combined results of the LEP Collaborations (1991)
at the Large Electron Positron Collider at CERN have shown
that there are only three generations of light neutrinos,
$N_\nu=3.00\pm 0.05~. $
Recent measurements of the neutron lifetime in a neutron bottle
and in a Penning trap resulted in the precise values,
$\tau_n=887.3\pm 3~ sec$ (Mampe et al. 1989) and
$\tau_n=893.6\pm 5.3~ sec$ (Byrne et al. 1990), respectively, yielding
a weighted ``world average value'' of (Hikasa et al. 1992) of
$\tau_n=889.1\pm 2.1~sec~ $
when combined with all previous measurements. Finally, the
precise measurements of the spectrum of the cosmic microwave
background radiation (MBR)
by the Far Infrared Absolute Spectrometer (FIRAS) on
the Cosmic Background Explorer (COBE) satellite (Mather et al. 1990)
and by a helium cooled spectrometer carried by
a high altitude rocket (Gush, Halpern and Wishnow 1990)
gave a black body temperature of $T=2.736\pm 0.017~K,$  which yields a
number density of relic photons from the Big Bang of
$n_\gamma=20.3T^3\approx 415\pm 8~cm^{-3}. $
As a result, the SBBN predictions for the primordial abundances
of the light elements depend essentially on a single unknown parameter,
$n_{B}~,$ the mean baryon number density in the Universe. Hence, the
primordial abundances of the light elements that are inferred
from observations, if correct, can be used both to test the SBBN
theory and to determine the mean baryon density in the Universe.
Indeed, in recent papers Olive et al. (1990) and WSSOK
have claimed to show that for baryon to photon ratio
            $2.8\leq \eta_{10}\leq 4~, $
where $\eta_{10}\equiv \eta\times 10^{10},$ the predictions of
the SBBN theory agree with the observations and that this
ratio implies (Schramm 1991)
that most of the nucleons in the Universe are dark.
However, in this paper we show that:

(a) The confidence level of the alleged agreement between
the predictions of SBBN theory and the primordial
abundances of the light elements that were inferred from observations
by WSSOK is less than 0.02\% for any value of $\eta~.$

(b) If the highly uncertain upper bound on primordial
D+$^3$He that was estimated by WSSOK  is relaxed (see for instance
Delbourgo-Salvador, Audouze and Vidal-Madjar 1987; Audouze 1987)
and only the less restrictive, but more reliable, lower bounds
on primordial D and $^3$He (their
presolar abundances as inferred from meteorites) are trusted, then
agreement between the SBBN theory and the primordial
abundances of $^4$He, D, $^3$He and $^7$Li as inferred from observations
is achieved with a high confidence level ($\gsim 70\%$)
provided that $\eta_{10}\approx 1.60\pm 0.10~.$

(c) The above value for $\eta$ yields a
mean baryon mass density in the Universe which is
consistent with the best estimates of the total mass density of
luminous matter (in the V, UV, IR, X and Radio bands) in the Universe.
Thus, SBBN theory and the observed abundances of the light elements
do not provide any reliable evidence that most of the baryons in the
Universe are dark.

(d) Since there is considerable dynamical evidence
from clusters of galaxies and large scale structures that
the total mass-energy density in the Universe is  greater
than the mean density of luminous matter by more than an order of
magnitude (see for instance Kolb and Turner 1990;
Mushotzky 1991; Schramm 1991), it follows that most of the
matter in the Universe is non baryonic dark matter.
\bigskip
{\bf II. THE SBBN PREDICTED ABUNDANCES}

The primordial abundances of $^4$He, D, $^3$He
and $^7$Li that are
predicted by the SBBN theory
with $N_\nu=3$ and $\tau_n=889.1~s$
are displayed in Fig.1  for $1\leq\eta_{10}\leq 10~.$
These are essentially the numerical results of
WSSOK that were also obtained from the
analytical calculations of Dar and Rudzsky (1992)
using the same nuclear reactions cross sections. Over the range
$2\lsim \eta_{10}\lsim 8,$ both the numerical results of WSSOK
and the analytical calculations are well approximated
(maximal deviation smaller than the uncertainty due to the
uncertainties in the nuclear cross sections)
by simple interpolating formulae:

{\bf Helium 4;} The primordial mass
fraction of $^4$He is well described by
$${\rm Y_p= 0.228+0.0105ln\eta_{10}+0.000208(\tau_n-889.1~s)}.
\eqno\eq $$
{\bf Deuterium;} The primordial abundance of D,
${\rm y_{2p}\equiv N(D)/N(H)}~,$
where N(X) is the number density of atoms X, is well described by
$${\rm y_{2p}= 4.6\times 10^{-4}\eta_{10}^{-1.67}}.
\eqno\eq $$
{\bf Helium 3;} The primordial abundance of $^3$He,
${\rm y_{3p}\equiv N(^3He)/N(H)}~,$
is well described by
$${\rm y_{3p}= 3.0\times 10^{-5}\eta_{10}^{-0.50}}.
\eqno\eq $$
{\bf Lithium 7}; The primordial abundance of $^7$Li,
${\rm y_{7p}\equiv N(^7Li)/N(H)}~,$
is well described by
$${\rm y_{7p}= 5.2\times 10^{-10}\eta_{10}^{-2.43}
+6.3\times 10^{-12}\eta_{10}^{2.43}}.
\eqno\eq $$
The uncertainties in the absolute normalizations of the predicted
abundances of $^4$He, D, $^3$He and $^7$Li due to uncertainties
in nuclear cross sections are,
$\pm 0.2\%,~\pm 4\%,$
$\pm 6\%$ and $\pm 20\%,$ respectively.
\bigskip
{\bf III. INFERRED PRIMORDIAL ABUNDANCES}

{\bf Helium 4:} The primordial mass fraction ${\rm Y_p}$ has been
determined from observations of the most metal-poor
extragalactic H II regions (see for instance Pagel 1990 and references
therein) by a linear
fit of ${\rm Y_p}$ versus metalicity, using O, N and C as
metalicity indicators.
The extrapolations of these results
to zero metalicity
by WSSOK gave
  $${\rm O:~~  Y_p=0.229\pm 0.004;~~
         N:~~  Y_p=0.231\pm 0.003;~~
         C:~~  Y_p=0.230\pm 0.007,   }\eqno\eq $$
where the errors are one standard deviations. Fuller, Boyd and Kalen
(1991), however, have argued that
the straight line fit to the $^4$He data to obtain the zero metalicity
intercept is inappropriate and places inordinate weight on high
metalicity points. Using a linear regression procedure (Lyons 1986)
for extrapolating lower metalicity data
to zero metalicity, and
N and O as metalicity indicators, FBK found
  $${\rm O:~~  Y_p=0.223\pm 0.009;~~
         N:~~  Y_p=0.220\pm 0.010.~~
           }\eqno\eq$$
{\bf Deuterium:}
Deuterium is easily destroyed already at relatively low temperatures.
Consequently, its local abundance depends strongly
on local chemical evolution and can vary markedly from site to site.
Since the sun burned D into $^3$He
during its pre-main sequence evolution, the difference between the
largest observed $^3$He abundance in gas rich meteorites and the
smallest observed $^3$He abundance in carbonaceous chondrites
meteorites was used by WSSOK to estimate the presolar D abundance and
to set a lower bound on primordial D due to its possible
destruction during presolar galactic chemical evolution:
$$ {\rm y_{2p}\geq   1.8\times 10^{-5}}. \eqno\eq $$
The local chemical evolution that preceded the solar system
is highly uncertain and precludes any reliable local
determination of primordial D.

{\bf Helium 3:} The smallest observed $^3$He
abundance in carbonaceous chondrites meteorites was used
by WSSOK as an estimate of its presolar abundance and as
a lower bound on primordial $^3$He (due to possible
destruction during presolar chemical evolution):
$$ {\rm y_{3p}}\geq 1.3\times 10^{-5}. \eqno\eq $$
WSSOK have also used a model dependent estimate for the survival
of $^3$He during local presolar chemical evolution
(primordial and that produced by burning away D)
to bound (2$\sigma$) primordial D+$^3$He:
$$3.3\times 10^{-5}\leq{\rm y_{23p}}\leq 1\times 10^{-4}. \eqno\eq $$
{\bf Lithium 7:} The primordial abundance of $^7$Li ,
was determined from the most metal poor, Population II halo stars.
Such stars, if sufficiently warm $(T\gsim 5500K)$,
have apparently not depleted their surface Lithium and are expected
to have nearly a constant Lithium abundance reflecting the Lithium
abundance present at the early evolution of the Galaxy (Spite and
Spite 1982a,b). WSSOK have fitted a plateau value
for 35 such stars with $T\geq 5500K~,$ which were selected from several
observations, and concluded that
$$ {\rm y_{7p}=(1.20\pm 0.21)\times 10^{-10}}. \eqno\eq $$
Deliyannis and Demarque (1991), however,
in their most recent paper derived from their detailed study
of Population II halo stars with $5500K\lsim T \lsim 6400K$
a more conservative estimate for the early Galaxy Lithium abundance,
$$ {\rm y_{7p}=(1.58\pm 0.35)\times 10^{-10}}. \eqno\eq $$
\bigskip
{\bf IV. COMPARISON BETWEEN THEORY AND OBSERVATIONS}

In order to test the agreement between the
SBBN theory and  observations
we applied to them the standard $\chi^2$ test (we have used the standard
CERN Computer Program MINUIT, James and Ross 1989), assuming the
errors to be statistical. First the SBBN predictions for
the primordial abundances of $^4$He, D+$^3$He and $^7$Li
as obtained by WSSOK (Section II) were compared with the corresponding
primordial abundances that were inferred by WSSOK from observations
(Section III), for baryon to photon ratio in the range $1\leq\eta_{10}
\leq 10~.$ The confidence level of the ``agreement''
between theory and ``observations'' as a function of
 $\eta_{10}$ is displayed in Fig.1.
Indeed, the best ``agreement'' is obtained for $\eta_{10}\approx 3.1~$,
but its confidence level is extremely poor,  only 0.02\%
(errors were treated as purely statistical).
In other words, the probability to be wrong in stating that the SBBN
predictions and the primordial abundances of the light elements
(with their quoted errors) as inferred from observations by WSSOK,
disagree for any value of $\eta$, is less than 0.02\% ! Consequently,
the confidence level of the value of the baryon mass density
that was deduced by WSSOK, and of their conclusion that most of the
baryons in the Universe are dark, is less than 0.02\% !

The failure of the SBBN to reproduce the WSSOK primordial abundances
suggests that something is wrong either with the SBBN theory or with the
WSSOK abundances, or with both. However, the qualitative
agreement between the SBBN predictions and
the ``observed'' primordial abundances of the light
elements that range over ten orders of magnitude, as well as
other impressive successes of the Standard Big Bang Model
(for a recent review see Peebles et al. 1991) suggest that the
failure is probably due to the WSSOK primordial abundances. Perhaps
not all the values inferred by WSSOK for the primordial abundances
of the light elements are correct. In fact, various authors
(see for instance
Delbourgo-Salvador, Audouze and Vidal-Madjar 1987; Audouze 1987
and references therein) have questioned
the upper bound on the primordial abundance of D+$^3$He
that was inferred by WSSOK from observations
because of large uncertainties in our knowledge of the local presolar
chemical evolution. Indeed, even in the local interstellar
medium (LISM) D abundances which were determined specroscopically
from absorption spectra  of the LISM (see for instance Murthy 1991)
are very different for different directions. Thus, we have repeated
the standard $\chi^2$ test for the SBBN predictions for primordial
$^4$He, D, $^3$He and $^7$Li,
ignoring the uncertain WSSOK upper bound on primordial D+$^3$He.
Satisfactory agreement (confidence level higher than 70\%) between
theory and ``observations'' was obtained for
$$\eta_{10}\approx 1.60\pm 0.1\eqno\eq $$
when we used the WSSOK estimates of primordial  $^4$He (Eq.5),
the WSSOK lower bounds for primordial D and $^3$He (Eqs. 7,8)
and the estimate by Deliyannis and Demarque (1991) of
primordial $^7$Li (Eq. 11).
This is demonstrated in Fig.2. Essentially the same values of $\eta$
and confidence level were obtained when
we used the FBK estimates (Eq. 6) for the primordial abundance of
$^4$He instead of the WSSOK estimates.
The corresponding baryon number density is
$$ n_{B}=\eta n_{\gamma}= (6.6\pm 0.5)\times
10^{- 8}cm^{-3} \eqno\eq $$ and the
baryon mass density is $\rho_{B}\approx n_{B} m_{P} =(1.11\pm 0.08)\times
10^{-31}~ g~cm^{-3}.$ When expressed
in critical density units, $\rho_{C}\equiv 3H_0^2/8\pi G~\approx
1.88\times 10^{-29}h^2g~cm^{-3},$  this baryon mass density is
$$\Omega_B\equiv \rho_{B}/\rho_{C}=(0.0059\pm 0.0007)h^{-2}. \eqno\eq $$
The observed light density in the Universe was estimated
(e.g., Felten 1987) to be
$ (2.4\pm 0.4)\times 10^8L_\odot h~ Mpc^{-3}.$ The mass to light ratio
which best reproduces rotation curves
within the visible  part of spiral galaxies (e.g., Rubin 1991) and X-ray
luminosities from the visibile part of
elliptical galaxies is (Mushotzky 1991), $M/L\approx (7\pm 3)hM_\odot/
L_\odot~.$ Consequently, the mean mass density of luminous matter in
the Universe is estimated to be,
$$\Omega_{LUM}\approx 0.0060\pm 0.0027 . \eqno\eq $$
This mass density is not in disagreement with the mass density
of baryonic matter obtained above from SBBN theory, in
particular if one notes that the recent best estimates of the value
of the Hubble parameter yield
$0.75\leq h \leq 1.00 $ (Jacoby et al. 1990,
Fukugita and Hogan 1991, Tonry 1991).
Conversely, if one constrains baryon mass density in SBBN to
be equal to the observed density of luminous
matter and h to the range $0.75 \leq h \leq 1.00,$ one obtains that
$\eta_{10} = 1.57\pm .22$ and
$\Omega_{B}\approx \Omega_{LUM}\approx 0.59\pm 0.15  $
with a confidence level of 85\% . For such baryonic mass density the
SBBN theory predicts that the primordial abundances of D and $^3$He
are ${\rm y_{2p}=(2.10\pm 0.35)\times 10^{-4}}$
and ${\rm y_{3p}=(2.37\pm 0.18)\times
10^{-5},}$ respectively.
\bigskip
 {\bf V. CONCLUSIONS}

The confidence level of the alleged agreement between the primordial
abundances of the light elements that were inferred from observations
by WSSOK and the predictions of the SBBN theory
is less than 0.02\% for any value of $\eta$ (including
$2.8\leq \eta_{10}\leq 4$). Thus, SBBN theory
and the WSSOK abundances do not provide
any reliable evidence for the existence of cosmologically
significant quantities of baryonic dark matter.
Moreover, agreement between the SBBN theory and observations
is achieved with a confidence level higher than 70\% for baryon to
photon ratio $\eta_{10}=1.60\pm 0.22~,$ if the highly
uncertain essentially theoretical
upper bound for primordial D+$^3$He that was estimated by WSSOK
is ignored. This range of $\eta$ is essentially dictated by
the present best esimates of primordial $^4$He and
$^7$Li from observations and their uncertainties.
It yields a mean baryon mass density of $\Omega_{B}=
(0.0058\pm 0.0007)h^{-2},$ which is
consistent with the best estimates of the mass density of
luminous (baryonic) matter in the Universe,
$\Omega_{LUM}=0.0060\pm 0.0027~,$
provided that the most recent estimates
$0.75\leq h \leq 1.00 $ (Jacoby et al 1990,
Fukugita and Hogan 1991, Tonry 1991) are correct.
Since, there is dynamical evidence
from clusters of galaxies and large scale structures that the
total mass-energy density in the Universe satisfies
(see for instance Kolb and Turner 1990 and references therein)
$\Omega\gsim 0.15~,$ it implies that
$\Omega \gg \Omega_{B}~,$
i.e., that most of the gravitating
matter in the Universe is non baryonic dark matter.
Finally, SBBN with a baryon mass
density similar to that of luminous matter predicts
primordial abundances,
${\rm y_{2p}=(2.10\pm 0.35)\times 10^{-4}}$ and
${\rm y_{3p}=(2.37\pm 0.18)\times
10^{-5}},$ for D and $^3$He,
respectively. These predictions, in principle, can be tested
by measuring high redshift absorption line
systems in quasar spectra due to D and $^3$He in intergalactic
space, perhaps by the Hubble Space Telescope.
\bigskip
{\bf Acknowledgement:} This work was
supported in part by the USA-Israel
Binational Science Foundation and The Technion Fund For Promotion
of Research. M. R. was supported in part by the Israel Ministry
of Immigrant Absorption.
\bigskip
\centerline {{\bf REFERENCES}}
\bigskip
The LEP Collaborations, 1991, CERN preprint CERN-PPE/91-232.

Audouze, J., 1987, ``Observational Cosmology'', (Eds. A. Hewitt,
G. Burbidge and L.Z. Fang; Reidel 1987) p. 89.

Alpher, R.A., Follin, J.W. and Herman, R.C., 1953, Phys. Rev.
{\bf 92}, 1347.

Boesgard, M. and Steigman, G., 1985, Ann. Rev. Astron.
Astrophys. {\bf 23}, 319.

Byrne, J. et al., 1990, Phys. Rev. Lett. {\bf 65}, 289.

Dar, A. and Rudzsky, M., 1992, to be submitted to Ap. J.

Delbourgo-salvador, P., Audouze, J. and Vidal-Madjar, A., 1987,
A\&A {\bf 174}, 365.

Deliyannis, C.P. and Demarque, P., 1991, Ap. J. {\bf 379}, 216.

Felten, J.E., 1987, ``Dark Matter In The Universe'' (Eds. J.
Kormendy and G.R. Knapp; Reidel 1987) p. 111.

Fukugita, M. and Hogan, C.J., 1991, Ap. J. Lett. {\bf 368}, L11.

Fuller, G.M., Boyd, R.N. and Kalen, J.D., 1991, Ap. J. Lett. {\bf 371},
L11.

Gush, H.P., Halpern, M. and Wishnow, E.H., 1990, Phys. Rev. Lett.
{\bf 65}, 537.

Hayashi, C., 1950,  Prog. Theor. Phys. {\bf 5}, 224 \& 235.

Hikasa, H. et al. (Particle Data Group), 1992, Phys. Rev. {\bf D45}, 1.

Jacoby, G.H. et al., 1990, Ap. J. {\bf 356}, 332.

James, F. and  Roos, M., 1989,
CERN Program Library report D506/1989.12.01.

Kolb, R. and Turner, M., 1990, ``The Early Universe'' (Addison
Wesley-1990).

Lyons, L., 1986, ``Statistics For  Nuclear And Particle
Physics.'' (Cambridge Univ. Press, 1986), p. 137.

Mampe, W. et al., 1989, Phys. Rev. Lett. {\bf 63}, 593.

Mather, J.C. et al., 1990, Ap. J. Lett. {\bf 354}, L37.

Murthy, J. et al., 1991, Ap. J. {\bf 356}, 223; E-{\bf 378}, 455.

Mushotzky, R., 1991, A.I.P. {\bf 222}, 394.

Olive, K. et al., 1990, Phys. Lett. {\bf B236}, 454.

Olive, K., 1991, Nuc. Phys. B (Proc. Suppl.) {\bf 19}, 36.

Pagel, B. E. J., 1990, Nordita Preprint, Nordita 90/47 A.

Peebles, P.J.E., 1966, Ap. J. {\bf 146}, 542.

Peebles, P.J.E. et al., 1991, Nature  {\bf 352}, 769.

Rubin, V. C., 1991, A.I.P. {\bf 222}, 371.

Schramm, D.N., 1991, Fermilab Preprint Conf-91/179-A.

Schramm, D.N. and Wagoner, R.V., 1977, Ann. Rev. Nucl. Part. Sci.,
{\bf 27}, 37.

Spite, M. and Spite, F., 1982a, Nature {\bf 297}, 483.

Spite, M. and Spite, F., 1982b, A\&A {\bf 115}, 357.

Tonry, J.L., 1991, Ap. J. Lett. {\bf 373}, L1.

Walker, T.P., Steigman, G., Schramm, D.N., Olive, K.A. and Kang, H.S.,
1991, Ap. J. {\bf 376}, 51.

Wagoner, R.V., Fowler, W.A. and Hoyle, F., 1967, Ap. J. {\bf 148}, 3.

Wagoner, R.V., 1969, Ap. J. Suppl. {\bf 162}, 247.

Wagoner, R.V., 1973,  Ap. J. {\bf 179}, 343.

Weinberg, S., 1972 ``Gravitation And Cosmology'' (John Wiley, 1972).

Yang, J. et al., 1984, Ap. J. {\bf 281}, 493.

\bigskip
\centerline{{\bf FIGURE CAPTIONS}}

{\bf Figure 1.} (a) The predicted primordial mass fraction of $^4$He and
the abundances (by number) of D, $^3$He, D+$^3$He and $^7$Li
as a function of $\eta_{10}$. Also shown are the 95\% confidence level
bounds that were inferred from observations by WSSOK.
The vertical band delimits the alleged range of $\eta_{10}$ where
predictions agree with observations. (b) The
values of $\chi^2$ (left scale)
and the corresponding confidence level (right scale)
of the alleged agreement between the predicted abundances and those
inferred by WSSOK from observations, as function of $\eta_{10}~.$
Best ``agreement'' is obtained for $\eta_{10}\approx 3.1~,$
but its confidence level is only 0.02\%!

{\bf Figure 2.} (a) The predicted primordial mass fraction of $^4$He and
the abundances (by number) of D, $^3$He, D+$^3$He and $^7$Li
as a function of $\eta_{10}$. Also shown are the primordial
mass fractions of $^4$He ($1\sigma$ errors) that were inferred
by WSSOK from observations of extragalactic HII regions
by interpolating to zero metalicity, using O, N and C as metalicity
indicators, the primordial abundance of $^7$Li ($1\sigma$ error)
that was inferred by Deliyannis and Demarque (1991) from observations
of metal poor Population II halo stars and the lower bounds of WSSOK on
primordial D and $^3$He from meteorites.
The vertical line indicates the value of $\eta_{10}$ which is
most consistent
with the observations. (b) The values of $\chi^2$ (left scale)
and the corresponding confidence level (right scale)
of the agreement between the predicted abundances and those
inferred from observations, as function of $\eta_{10}~.$
Best agreement is obtained for $\eta_{10}\approx 1.60~$ with
a confidence level above 70\%.
\endpage
\end